\documentclass[10pt,a4paper,twocolumn,oneside,conference]{IEEEtran}

\usepackage{cite}
\usepackage{graphicx}
\usepackage{amsmath}
\usepackage{times}
\usepackage{latexsym}
\usepackage{graphicx}
\usepackage{bm}
\usepackage{amssymb}
\usepackage[center]{caption2}
\usepackage{stfloats}
\usepackage{cases}
\usepackage{array}
\usepackage{setspace}
\usepackage{fancyhdr}
\usepackage{algorithm}
\usepackage{algpseudocode}

\usepackage{amsmath,amsthm}
\usepackage{balance} %balance package 1
\usepackage{flushend} %balance package 2

\usepackage{graphicx}
\usepackage{subfigure}

\usepackage{xcolor}
\allowdisplaybreaks[4]

\newcaptionstyle{mystyle1}{%
  \centering TABLE  \captiontext \par}
\captionstyle{mystyle1}
\newcaptionstyle{mystyle2}{%
  \captionlabel $.$ \: \captiontext \par}
\captionstyle{mystyle2}
\newcaptionstyle{mystyle3}{%
  \captionlabel $.$ \: \captiontext \par}
\captionstyle{mystyle3}

\begin{document}

\title{{Energy Efficient Power Control for the Two-tier Networks with Small Cells and Massive MIMO}}

\author{
\IEEEauthorblockN{Ningning Lu$^\dag$, Yanxiang Jiang$^\dag$$^{*}$, Fuchun Zheng$^\ddag$, and Xiaohu You$^\dag$}
\IEEEauthorblockA{$^\dag$National Mobile Communications Research Laboratory,
Southeast University, Nanjing 210096, China.\\
$^\ddag$School of Systems Engineering, University of
Reading, Reading, RG6 6AY, UK.\\
$^*$E-mail: yxjiang@seu.edu.cn}
}

\maketitle

\begin{abstract}
In this paper, energy efficient power control for
the uplink two-tier networks where a macrocell tier with a massive multiple-input multiple-output (MIMO) base station is overlaid with a small cell tier is investigated.
We propose a distributed energy efficient power control algorithm which allows each user in the two-tier network taking individual decisions to optimize its own energy efficiency (EE) for the multi-user and multi-cell scenario.
The distributed power control algorithm is implemented by decoupling the EE optimization problem into two steps.
In the first step, we propose to assign the users on the same resource into the same group and each group can optimize its own EE, respectively.
In the second step, multiple power control games based on evolutionary game theory (EGT) are formulated for each group, which allows each user optimizing its own EE.
In the EGT-based power control games, each player selects a strategy giving a higher payoff than the average payoff, which can improve the fairness among the users.
The proposed algorithm has a linear complexity with respect to the number of subcarriers and the number of cells in comparison with the brute force approach which has an exponential complexity.
Simulation results show the remarkable improvements in terms of fairness by using the proposed algorithm.
\end{abstract}

%\begin{keywords}
%  Energy efficiency, fairness, power control, evolutionary game theory, two-tier networks.
%\end{keywords}

%\newpage
\section{Introduction}
  {As the new technologies in 5G mobile communication systems,small cells and massive multiple-input multiple-output (MIMO)\cite{Andrews} can greatly improve the spectral efficiency and energy efficiency (EE), which have received much attention in recent years.}
  Although small cells and massive MIMO have been compared in\cite{Liu1}, they are not necessarily rivals.
  Actually, a two-tier network architecture that incorporates the small cells and massive MIMO can attain the benefits of both technologies\cite{Hoydis,Li}.
  In the two-tier network architecture, massive MIMO ensures full-area coverage, while small cells mainly serve indoor and outdoor hotspots.
  {The power consumption of the two-tier network was characterized and analyzed in \cite{Sanguinetti}.
   A spatial interference coordination scheme was proposed to protect small cell users from macro-cell interference in \cite{Dommel}.
  Three low-complexity strategies for explicit inter-tier interference coordination through spatial blanking were developed in \cite{Adhikary}.}

  Although the two-tier network may provide capacity enhancements, it poses many challenges to the power control of the two-tier network.
  Firstly, due to the large number of small cells, centralized  power control algorithm may be infeasible for the self-organized small cells and low-complexity power control algorithm is needed.
  Secondly, there exists not only fairness problem between the macrocell and the small cells, but also fairness problem among the small cells in the two-tier network.
  Game theory is suitable to address the problem of power control in self-organizing small cells since it allows the players to learn from the environment and take individual decisions for attaining the equilibrium with minimum information exchange.

  Non-cooperative game theory (NGT) was applied to describe the system model and implement the distributed resource allocation in \cite{Eun,Liang,Hanjun1}.
  A distributed power control scheme for closed access femtocell networks was proposed in \cite{Eun} to minimize interference to macro users.
  A distributed coverage optimization algorithm using game theory was proposed in \cite{Liang} for the self optimization network architecture of small cell cluster.
  The joint uplink subchannel and power allocation problem in cognitive small cells using cooperative Nash bargaining game theory was investigated in \cite{Hanjun1}.

 {
  However, methods to guarantee the effectiveness of Nash equilibrium influencing the fairness among the users still need to be investigated.
  Thus, the power control algorithm to improve the fairness among the users is also needed in the two-tier networks.}
  In \cite{Hanjun2}, the authors proposed a resource allocation scheme to maximize the total capacity and improve the fairness.
  However, the power control algorithms in \cite{Eun,Liang,Hanjun1,Hanjun2} were proposed for small cells underlaying traditional macro cellular networks.
   {There is little work about  power control on the two-tier networks where a macrocell tier with a massive MIMO base station is overlaid with a small cell tier.}

  {Motivated by the aforementioned results, we propose a distributed energy efficient power control algorithm for the uplink two-tier networks with small cells and massive MIMO.
  The distributed power control algorithm is implemented by decoupling the EE optimization problem into two steps for the multi-user and multi-cell scenario.
 In the first step, we propose to assign the users into different groups.
  In the second step, multiple power control games based on evolutionary game theory (EGT) are formulated for each group.
  Then, each user in the two-tier networks can take individual decisions to optimize its own energy efficiency.
  Thus, the computational complexity can be greatly reduced in comparison with the brute force approach and the fairness among the users can be improved significantly.
}

  The rest of this paper is organized as follows.
  In Section II, the system model of the two-tier  { networks   with} small cells and massive MIMO is presented.
  In Section III, the distributed energy efficient power control algorithm is proposed. Simulation results are shown in Section IV, and final conclusions are drawn in Section V.

%\pubidadjcol
\section{System Model}
\begin{figure}[!t]
\centering
\includegraphics[width=0.45\textwidth]{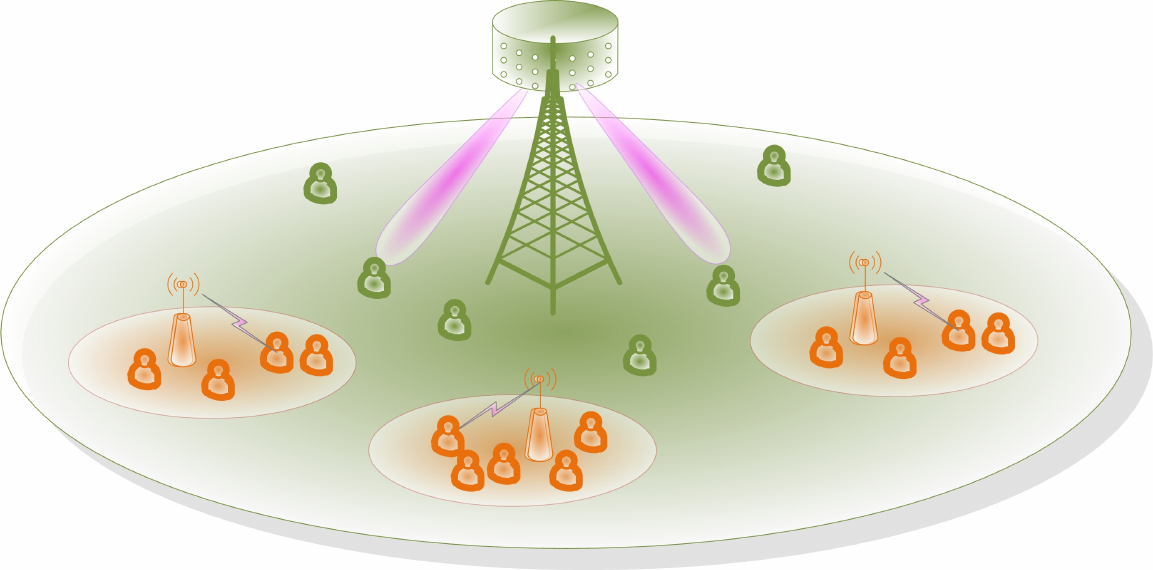}
\captionstyle{mystyle3}
\caption{System model of the two-tier network.}
\label{smallcell_massiveMIMO}
\end{figure}

{In this paper, the two-tier networks, where a macrocell tier with a massive MIMO base station is overlaid with a tier of small cells, are considered.
We investigate the uplink transmission based on the orthogonal frequency division multiple access (OFDMA) operation.
The two-tier networks }are  illustrated in Fig. \ref{smallcell_massiveMIMO}, which includes a macro base station (MBS) with \({N_\text{MBS}}\)  antennas and $K$ SBSs with \({N_\text{SBS}}\) antennas.
{We assume} that  \({N_\text{MBS}} \gg {N_\text{SBS}}\).
The small cells share the same set of orthogonal subcarriers with the macrocell and the number of the subcarriers is assumed to be $N$.
{We suppose }that the number of users served by the MBS is \({N_0}\), and the number of users served by the $k$-th SBS is \({N_k} (k = 1,2,...,K)\).
Each user selects a transmit power level from a finite set of values denoted by \(\mathcal{L}=\{ 1,2,...,L\} \).

The impulse responses of all the channels are assumed to be flat fading.
The received signal at the MBS can be expressed as:
\begin{equation}\begin{split}\label{Macro SIN}
{\boldsymbol{y}_0}{\bf{ = }}{{\boldsymbol{G}}_{00}}{{\boldsymbol{p}}_0}^{1/2}{{\boldsymbol{x}}_0} + \sum\limits_{k = 1}^K {{{\boldsymbol{G}}_{k0}}{{\boldsymbol{p}}_k}^{1/2}{{\boldsymbol{x}}_k}} { +\boldsymbol{ n}}
\end{split}\end{equation}
where
${{\boldsymbol{G}}_{00}} = {{\boldsymbol{H}}_{00}}{{\boldsymbol{D}}_{00}}^{1/2} \in {\mathcal{C}^{{N_\text{MBS}} \times {N_0}}}$ is the channel coefficient matrix between the MBS and its users,
${{\boldsymbol{G}}_{k0}} = {{\boldsymbol{H}}_{k0}}{{\boldsymbol{D}}_{k0}}^{1/2} \in {\mathcal{C}^{{N_\text{MBS}} \times {N_k}}}$ is the channel coefficient matrix between the MBS and the users served by $k$-th SBS,
${{\boldsymbol{H}}_{00}} $ and ${{\boldsymbol{H}}_{k0}} \in {\mathcal{C}^{{N_\text{MBS}} \times {N_0}}}$ are the realizations of fast fading channels and their components are always assumed to be i.i.d. Rayleigh flat-fading random variables $\mathcal{N}(0,1)$,
 ${{\boldsymbol{D}}_{00}}$ and ${{\boldsymbol{D}}_{k0}} \in {\mathcal{C}^{{N_0} \times {N_0}}}$ are the diagonal matrices whose elements represent large scale fading factors between the MBS and the users,
${{\boldsymbol{p}}_0}$ and ${\boldsymbol{x}}_0^{}$ are the transmit power and the transmit data symbol of the users served by the MBS, respectively,
 ${{\boldsymbol{p}}_k}$  and ${\boldsymbol{x}}_k^{}$  are the transmit power and the transmit data symbol of the users served by the $k$-th SBS, respectively,
 ${\boldsymbol{n}}$  is the additive white Gaussian noise.

The received signal at the $k$-th SBS can be expressed as:
\begin{multline}\label{Smallcell SIN}
{{\boldsymbol{y}}_k}{\boldsymbol{ = }} \\
{{\boldsymbol{G}}_{kk}}{{\boldsymbol{p}}_k}^{1/2}{{\boldsymbol{x}}_k}{ +\boldsymbol{ G}}_{0k}^{}{{\boldsymbol{p}}_0}^{1/2}{{\boldsymbol{x}}_0}{ +\boldsymbol{ }}{\sum\limits_{l = 1, l \ne k}^{K}} {{{\boldsymbol{G}}_{lk}}{{\boldsymbol{p}}_l}^{1/2}{{\boldsymbol{x}}_l}} { +\boldsymbol{ n}}
\end{multline}
where ${{\boldsymbol{G}}_{kk}} = {{\boldsymbol{H}}_{kk}}{{\boldsymbol{D}}_{kk}}^{1/2} \in {\mathcal{C}^{{N_\text{SBS}} \times {N_k}}}$ is the channel coefficient matrix between the $k$-th SBS and  its users,
 ${{\boldsymbol{G}}_{0k}} = {{\boldsymbol{H}}_{0k}}{{\boldsymbol{D}}_{0k}}^{1/2} \in {\mathcal{C}^{{N_\text{SBS}} \times {N_0}}}$ is the channel coefficient matrices between the $k$-th SBS and the users served by the MBS,
 ${{\boldsymbol{G}}_{lk}} = {{\boldsymbol{H}}_{lk}}{{\boldsymbol{D}}_{lk}}^{1/2} \in {\mathcal{C}^{{N_\text{SBS}} \times {N_l}}}$ is the channel coefficient matrix between the $k$-th SBS and the users served by the $l$-th SBS,
 ${{\boldsymbol{H}}_{kk}} $, ${{\boldsymbol{H}}_{0k}}$ and ${{\boldsymbol{H}}_{lk}}\in {\mathcal{C}^{{N_\text{SBS}} \times {N_k}}}$ are the realizations of fast fading channels and their components are always assumed to be i.i.d. Rayleigh flat-fading random variables $\mathcal{N}(0,1)$,
 ${{\boldsymbol{D}}_{kk}}$, ${{\boldsymbol{D}}_{0k}}$ and ${{\boldsymbol{D}}_{lk}} \in {\mathcal{C}^{{N_k}\times {N_k}}}$ are the diagonal matrix whose elements represent large scale fading factors between the $k$-th SBS and the users,
 ${{\boldsymbol{p}}_l}$  and ${\boldsymbol{x}}_l^{}$  are the transmit power and the transmit data symbol of the users served by the $l$-th SBS, respectively,
The diagonal components of ${{\boldsymbol{D}}_{00}}$, ${{\boldsymbol{D}}_{l0}}$, ${{\boldsymbol{D}}_{kk}}$, ${{\boldsymbol{D}}_{0k}}$ and ${{\boldsymbol{D}}_{lk}}$ have the form of \({\beta } = \varphi \varsigma /d^\alpha \), where \(\varphi \) is a constant related to carrier frequency and antenna gain, \({d}\) is the distance between the BS and the corresponding user, \(\alpha \) is the path loss exponent, \(\varsigma \) represents the shadow fading with logarithmic normal distribution \(10{\log _{10}}\varsigma  \sim \mathcal{CN}(0,{\sigma ^2})\).

 {We assume} that the maximum-ratio combining (MRC) received matrices are adopted at the MBS and SBSs.
 Then, they can be written as:
\begin{equation}\begin{split}\label{MRC MBS}
{\boldsymbol{A}}_{kk}^{} = \frac{{{\boldsymbol{G}}_{kk}^{}}}{{\left\| {{\boldsymbol{G}}_{kk}^{}} \right\|}},k=0,1,...,K
\end{split}\end{equation}
where $\left\|  \cdot  \right\|$ represents the $L2$-norm.
 The received signal at the MBS or SBSs after the MRC received matrices can be written as:
\begin{equation}\begin{split}\label{MRC SBS}
{{\boldsymbol{z}}_k}={\boldsymbol{A}}_{kk}^{T}{{\boldsymbol{y}}_k},k=0,1,...,K
\end{split}\end{equation}

Correspondingly, the received  signal to interference and noise ratio (SINR) of the user served by the MBS on the $i$-th subcarrier can be written as:
\begin{equation}\begin{split}\label{Macro SINR}
\gamma_{0i}^{} = \frac{{p_{0i}^{}{{\left| {{\boldsymbol{a}}_{00i}^H{\boldsymbol{g}}_{00i}^{}} \right|}^2}}}{{\sum\limits_{l = 1}^K {p_{li}^{}{{\left| {{\boldsymbol{a}}_{00i}^H{\boldsymbol{g}}_{l0i}^{}} \right|}^2} + {{\left\| {{\boldsymbol{a}}_{00i}^{}} \right\|}^2}{N_0}} }}
\end{split}\end{equation}
and the received SINR of the user served by the $k$-th SBS on the $i$-th subcarrier can be written as:
\begin{multline}\label{Smallcell SINR}
\gamma_{ki}^{} = \frac{{p_{ki}^{}{{\left| {{\boldsymbol{a}}_{kki}^H{\boldsymbol{g}}_{kki}^{}} \right|}^2}}}{{p_{0i}^{}{\left| {{\boldsymbol{a}}_{kki}^H{\boldsymbol{g}}_{0ki}^{}} \right|}^2}+{\underset{l \ne k}{\sum\limits_{l = 1}^{K}} {p_{li}^{}{{\left| {{\boldsymbol{a}}_{kki}^H{\bf{g}}_{lki}^{}} \right|}^2} + {{\left\| {{\boldsymbol{a}}_{kki}^{}} \right\|}^2}{N_0}} }},
\\
k=1,2,...,K
\end{multline}
where $p_{0i}^{}$ and $p_{ki}^{}$ are the $i$-th diagonal elements of ${{\boldsymbol{p}}_0}$ and ${{\boldsymbol{p}}_k}$,
${\boldsymbol{a}}_{00i}^{}$ and ${\boldsymbol{a}}_{kki}^{}$ are the $i$-th column vectors of ${\boldsymbol{A}}_{00}^{}$ and ${\boldsymbol{A}}_{kk}^{}$,
${\boldsymbol{g}}_{00i}^{}$, ${\boldsymbol{g}}_{l0i}^{}$, ${\boldsymbol{g}}_{kki}^{}$, ${\boldsymbol{g}}_{0ki}^{}$ and ${\boldsymbol{g}}_{lki}^{}$ are the $i$-th column vectors of ${\boldsymbol{G}}_{00}^{}$, ${\boldsymbol{G}}_{l0}^{}$, ${\boldsymbol{G}}_{kk}^{}$, ${\boldsymbol{G}}_{0k}^{}$ and ${\boldsymbol{G}}_{lk}^{}$, respectively.

The data transmit rate of the user on the $i$-th subcarrier served by the MBS or the $k$-th SBS can be expressed as
\begin{equation}\begin{split}\label{Smallcell R}
r_{ki}^{} = {\log _2}(1 + \gamma_{ki}^{}), k = 0,1,...,K
\end{split}\end{equation}

The power consumption model of the user on the $i$-th subcarrier served by the MBS or the $k$-th SBS can be expressed as
\begin{equation}\begin{split}\label{Smallcell Psum}
P_{\text{sum},ki}^{} = p_{ki}^{} + p_{c,ki}^{}, k=0,1,...,K
\end{split}\end{equation}
where $p_{c,ki}$ is the static circuit power at user equipment side.
$p_{ki}^{}$ is set to be zero if there is no user on the $i$-th subcarrier served by the MBS or the $k$-th SBS.

\newtheorem{lemma}{Lemma}
\newtheorem{theorem}{Theorem}
With the rapid deployment of wireless communication technology, energy consumption has increased dramatically, which makes it imperative to achieve higher EE.
We define the EE of the user on the $i$-th subcarrier served by the MBS or the $k$-th SBS  as follows,
\begin{equation}\begin{split}\label{System EE}
{\rm{EE}}_k^{i} ={\frac{{r_{ki}}}{P_{\text{sum},ki}^{}} },k=0,1,...,K
\end{split}\end{equation}
The EE of the two-tier network based on \cite{Jiang} can be defined as
\begin{equation}\begin{split}\label{Overal EE}
{\rm{EE}}{\bf{ = }} \sum\limits_{k = 0}^K \sum\limits_{i = 1}^N{\rm{EE}}_k^{i}
\end{split}\end{equation}
This definition is based on the sum EE of all the users in the two-tier network rather than the ratio of the sum network throughput to the sum network power consumption.
This is because  neither powers of different users nor their EE in the two-tier network can be shared.
 Therefore, we focus on the EE of the users in uplink two-tier networks and the overall EE is optimized on condition that the EE of each user is optimal.

\section{The distributed power control algorithm}
{The SBSs can be randomly deployed either by the operators or by the users in the two-tier networks with small cells and massive MIMO.
In addition, the computational complexity of the EE optimization increases with the number of the subcarriers and the number of cells  for the multi-user and multi-cell scenario.
Thus, the centralized power control algorithm is not appropriate for the self-organized small cells, and distributed power control algorithm is needed in the two-tier networks.
Therefore, we propose a distributed energy efficient power control algorithm in this section to decrease the computational complexity greatly.
}

\subsection{ Formulation of the Distributed Energy Efficient Power Control Algorithm}
{The distributed power control algorithm is implemented by decoupling the EE optimization into two steps.
In the first step, we propose to decompose the process of the energy efficient power control. In the second step, multiple power control games based on EGT are formulated.
}
\subsubsection{ Decomposing of the Energy Efficient Power Control}
A brute force approach to solve the optimization problem related to (\ref{Overal EE}) has an exponential complexity with respect to the
number of subcarriers and the number of cells in the two-tier
network in the two-tier network.
{
Therefore, we propose to decompose the process of the energy efficient power control by dividing the users into groups.
The users on the same subcarrier are assigned to the same group.
Thus, the users can be divided into $N$ groups $\{g_1,g_2,...,{g_N}\}$, and the users on the $i$-th subcarrier are in the group $g_i\in \{g_1,g_2,...,{g_N}\}$.
Then, the original optimization problem can be decomposed into $N$ independent optimization subproblems, and each group can optimize its own EE, respectively.
The EE of the group $g_i$ can be expressed as
\begin{equation}\begin{split}\label{Group EE}
{\rm{EE}}_i{\bf{ = }} \sum\limits_{k = 0}^K {\rm{EE}}_k^{i}
\end{split}\end{equation}
%{\color{red}It can be observed that the complexity of the EE optimization can be reduced to linear complexity with respect to the number of subcarriers.}
}

\subsubsection{ Formulation of the Energy Efficient Power Control Games based on EGT}
{
On the one hand, the direct method to solve the optimization problem related to (\ref{Group EE}) has an exponential complexity with respect to the number of cells.
To further reduce the complexity, we propose a distributed power control scheme from a game theory perspective.
Thus, each user can optimize its own EE as defined in (\ref{System EE}), respectively.
%{\color{red}The complexity of the EE optimization can be reduced to linear complexity with respect to the number of cells.}
}

On the other hand, the fairness problem among the cells in the two-tier networks is concerned with the widespread deployment of the SBSs.
The fairness problem exists not only between the macrocell and the small cells, but also   among the small cells.
Furthermore, the performance of some users may be very poor by using the traditional game theory, which exacerbates the fairness problem.
Therefore, we propose to formulate the power control game by using EGT \cite{Niy}.
In the EGT-based power control games, each player selects a strategy which gives a higher payoff than the average payoff.
Thus, the proposed distributed power control algorithm can improve the fairness among the users in the two-tier networks.

We formulate $N$ power control games $\{G_1,G_2,...,{G_N}\}$ for the groups $\{g_1,g_2,...,{g_N}\}$ based on EGT.
The users on the same subcarrier are assigned to the same power control game.
For each game ${G_i} \in\{G_1,G_2,...,{G_N}\}$, the EGT-based power control game includes four main factors as follows.

Players: For the EGT-based power control game ${G_i}$, its players are all the users on the $i$-th subcarrier. And the number of the players in the game ${G_i}$ is $K+1$.

Actions: We define the action set for each player as
${A_i} = \{ {a_{i,1}},{a_{i,2}},...,{a_{i,L}}\}$ which includes all the possible transmit power strategies for each player.
Each player selects a suitable transmit power strategy from the strategy set ${A}$.

Population: In the context of the EGT-based power control game, the set of players also constitutes the population.
We define $k_{i,a}$ as the number of users selecting strategy $a \in {A_i}$ in the game ${G_i}$.
Then, the proportion of the population in the game ${G_i}$ choosing action $a$ is given by
\begin{equation}\begin{split}\label{SINR}
{x_{i,a}} = \frac{{{k_{i,a}}}}{K+1}
\end{split}\end{equation}
We can see that $\sum_{a \in {A}}x_{i,a}=1$.

{Payoff function: The payoff of each player is determined by its EE.
Let $\mathcal{K}_{i,a}$ denote the set of the users in the game ${G_i}$ selecting the strategy $a$, and
$\pi_{i,a}$  the payoff function of each player.
Then, $\pi_{i,a}$ can be expressed as
\begin{equation}\begin{split}\label{payoff}
{ \pi_{i,a}} = \frac{\sum\nolimits_{k\in\mathcal{K}_{i,a}}{\rm{EE}}_k^{i}}{(K+1){x_{i,a}}}
\end{split}\end{equation}
}

\subsubsection{ The Distributed Power Control Algorithm}
%In the multiple power control games, the payoff of the users in the two-tier network is maintained by a centralized controller.
%We primarily assign the users on the same subcarrier into the same group.
%Then, we formulate a power control game for each group.
The proposed distributed power control algorithm is presented in Algorithm 1.
{The distributed power control algorithm is implemented in two steps.
In the first step, we decompose the process of the energy efficient power control by dividing the users into $N$ groups.
The users on the same subcarrier are in the same group.
In the second step, $N$ power control games are formulated for each group based on EGT.}
For each power control game ${G_i} \in\{G_1,G_2,...,{G_N}\}$, all players initially play a randomly selected strategy.
Then, the players compare their payoffs with the average payoff of the population and select a strategy which would give a higher payoff than the average one.
If there is no transmit power strategy making its payoff higher than the average one, the user's strategy remains unchanged.
{
Then, the rest users continue to select the proper strategies in accordance with the above method.
If the strategies of all the users remain unchanged, the power control process will be terminated and the proposed algorithm will be convergent.
A sub-optimal solution to the maximization of EE can be attained by the proposed algorithm, while the fairness among the users can be greatly improved.
In addition, the proposed algorithm has a linear complexity with respect to the number of subcarriers and the number of cells in comparison with the brute force approach which has an exponential complexity.
}

\begin{algorithm}[!t]
\doublespacing
\caption{The Distributed Power Control Algorithm}
{\renewcommand\baselinestretch{1}\selectfont
\begin{itemize}
\item [1)] All the users in the two-tier network randomly choose an initialized transmit power strategy.
\item [2)] Assign the users to $N$ groups , and the users on the same subcarrier are in the same group.
\item [3)] Formulate $N$ power control games $\{G_1,G_2,...,{G_N}\}$ for each group.
\item [4)]  For each game $G_i$, \textbf{loop}
\begin{itemize}
    \item [5)] The user in the game $G_i$ computes the payoff ${\pi_{ki}}$ and sends it to the centralized controller.

    \item [6)] The centralized controller computes the average payoff
           \[\bar \pi_i  = \sum\limits_{k = 0}^K {{\pi _{ki}}}/({K+1})\]
        and broadcasts it back to the users.

    \item [7)] \textbf{if} ${\pi_{ki}}\leq\bar \pi_i$, \textbf{then}
    \begin{itemize}
          \item [8)] \textbf{if} the selected transmit power strategy is not the last one, \textbf{then}
          \begin{itemize}
                 \item [9)]the user selects another transmit power strategy.
          \end{itemize}
          \item [10] \textbf{else}
          \begin{itemize}
                 \item [11)]the transmit power strategy remains unchanged.
          \end{itemize}
          \item [12)] \textbf{end if}
    \end{itemize}

    \item [13)] \textbf{end if}
\end{itemize}
\item [14)] \textbf{end loop} for the users in the game $G_i$.
\end{itemize}
\par}
\end{algorithm}

\subsection{ Replicator Dynamics and Evolutionary Equilibrium}
In the considered EGT-based power control game, the strategy adaptation process and population state evolution of the players can be modeled by using a set of ordinary differential equations called replicator dynamics. For each games ${G_i} \in\{G_1,G_2,...,{G_N}\}$, the replicator dynamics can be defined as follows:
\begin{equation}\begin{split}\label{replicator dynamics}
{\dot x_{i,a}} = {x_{i,a}}({\pi _{i,a}} - {\bar \pi_{i}})
\end{split}\end{equation}
where ${\pi _{i,a}}$ is the payoff of the individuals in the game ${G_i}$ choosing strategy $a$ and the replicator dynamics satisfy the condition of $\sum_{a}\dot x_{i,a}=0$,
{ and ${\bar \pi_i}$ is the average payoff of the entire population, which can be expressed as}
\begin{equation}\begin{split}\label{entier average payoff}
\bar \pi_i  = \sum\limits_{a \in A} {{\pi _{i,a}}{x_{i,a}}}
\end{split}\end{equation}
 For the EGT-based power control game ${G_i}$, the average payoff is given by
\begin{equation}\begin{split}\label{game average payoff}
\bar \pi_i  = \sum\limits_{k = 0}^K {{\pi _{i,k}}}/({K+1})
\end{split}\end{equation}
The replicator dynamics are important for the evolutionary game since they can provide information about the population.
It is also useful to investigate the speed of convergence of strategy adaptation to attain the solution to the game.
Based on this replicator dynamics of the players in the two-tier network, the number of players choosing strategy $a$ increases if their payoff is above the average payoff.

We consider the evolutionary equilibrium as the solution to the EGT-based power control games, which is defined as the fixed points of the replicator dynamics.
We can obtain the equilibrium of the EGT-based power control games through population evolution.
The speed of strategy adaptation of each power control game is zero (i.e., $\dot x_{i,a}=0$) at the evolutionary equilibrium, where no player deviates to gain a higher payoff.
{
To evaluate the stability of the evolutionary equilibrium,
the eigenvalues of the Jacobian matrix corresponding to the
replicator dynamics need to be evaluated. The evolutionary
equilibrium in the proposed power control algorithm is stable
if all the eigenvalues have a negative real part \cite{Kuz}.
}

\subsection{ Complexity Analysis}
In this section, we compare the complexity of the proposed distributed power control algorithm with the centralized power control algorithm.
It can be seen that there are $L^{(K+1)N}$ strategies to solve the centralized optimization problem with respect to (\ref{Overal EE}).
We can describe the computational complexity as $\mathcal{O}(L^{(K+1)N})$.
{We can optimize the EE of the users on the same subcarrier individually after assigning the users on the same resource to the same group.
}
The computational complexity is $\mathcal{O}(L^{(K+1)})$ to solve the optimization problem with respect to (\ref{Group EE}).
Then, the computational complexity can be reduced to $\mathcal{O}(NL^{(K+1)})$ to solve the corresponding optimization problem.
{In order to further decrease the computational complexity, we propose to formulate multiple power control games for each group.
}
Thus, each user can optimize its own EE, respectively.
For each game ${G_i} \in\{G_1,G_2,...,{G_N}\}$, its players are all the users on the $i$-th subcarrier.
The action set for each player in the game ${G_i}$ can be described as
${A_i} = \{ {a_{i,1}},{a_{i,2}},...,{a_{i,L}}\}$ which includes all the possible transmit power strategies for each user.
Then, the computational complexity of the game $G_i$ is $\mathcal{O}((K+1)L)$. Correspondingly, the computational complexity of the distributed power control algorithm is $\mathcal{O}((K+1)LN)$.
It can be seen that the proposed algorithm has a
linear complexity with respect to the number of subcarriers and
the number of cells in comparison with the brute force approach
which has an exponential complexity.
Note here that lower complexity will contribute further to the reduction of energy consumption in baseband processing.
\section{Simulation Results}
In this section, the performance of the proposed distributed power control algorithm is evaluated via simulations.
In the simulations, we consider a two-tier networks including an MBS with 128 antennas and several SBSs with 4 antennas each.
Assume that the number of the users in each cell is the same for the sake of simplicity.
Let $N_u$ denote the number of the users in each cell.
The simulation parameters are shown in Table I.

\begin{table}[!t]
\centering
\caption{Simulation parameters}
\begin{tabular}{|c|c|}
\hline
Parameter & Value\\
\hline
%Cellular layout & Isolated cell\\
%\hline
The radius of the macrocell & 1000m\\
\hline
The radius of the small cells & 100m\\
\hline
Path-loss exponent $\alpha$ & 3.8\\
\hline
Noise spectral density  & -194dBm/Hz\\
\hline
Antennas number of the MBS, \({N_\text{MBS}}\) & 128\\
\hline
Antennas number of the SBS, \({N_\text{SBS}}\) & 4\\
\hline
Variance of log-normal shadow fading $\sigma ^2$& 10dB\\
\hline
Factor $\varphi$& 1\\
\hline
Constant power per user \(p_{c,ki}^{}\) &0.01W\\
\hline
\end{tabular}
\end{table}

\begin{figure}[!t]
\centering %\vspace*{135pt}
\includegraphics[width=0.44\textwidth]{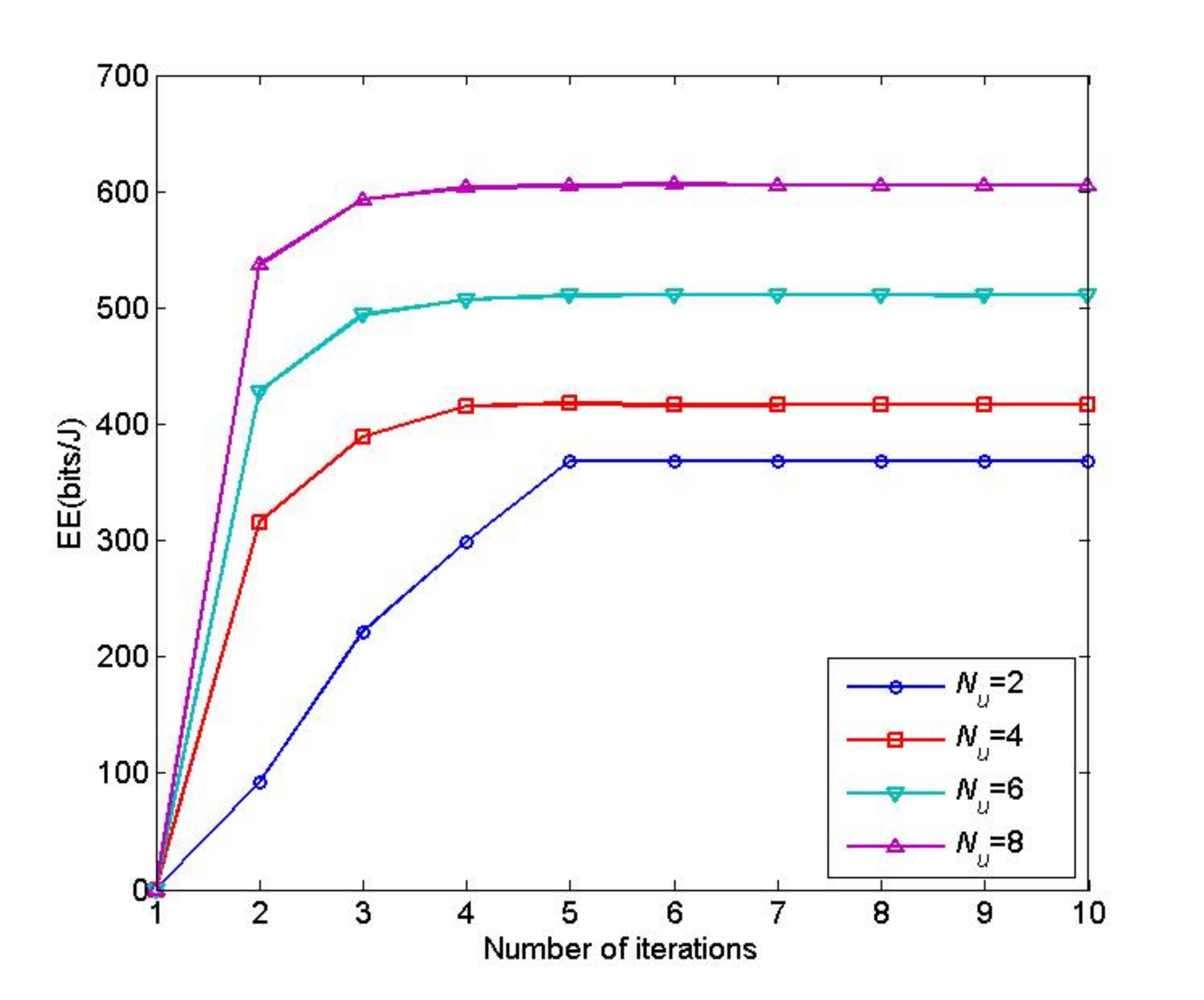}
%\captionstyle{mystyle3}
\caption{EE versus the number of iterations with \(K=2\).}% and different $N_u$.}
\label{fig4}
\end{figure}

In Fig. \ref{fig4}, we show the evolution of the proposed distributed algorithm with \(K=2\) and different $N_u$.
The EE in the figure is the average payoff.
It can be observed that the proposed distributed algorithm converges within several iterations for all the considered \(N_u\).
It can also be observed from the figure that the EE increases with $N_u$ by using the proposed EGT-based algorithm.
The reason is that different users in the same cell do not crosstalk each other and the users can be allocated the power that are more suitable for them.
Then, multiuser diversity can be exploited.

\begin{figure}[!t]
\centering %\vspace*{135pt}
\subfigure[The EGT-based algorithm]{\label{1111}
\includegraphics[width=0.2\textwidth]{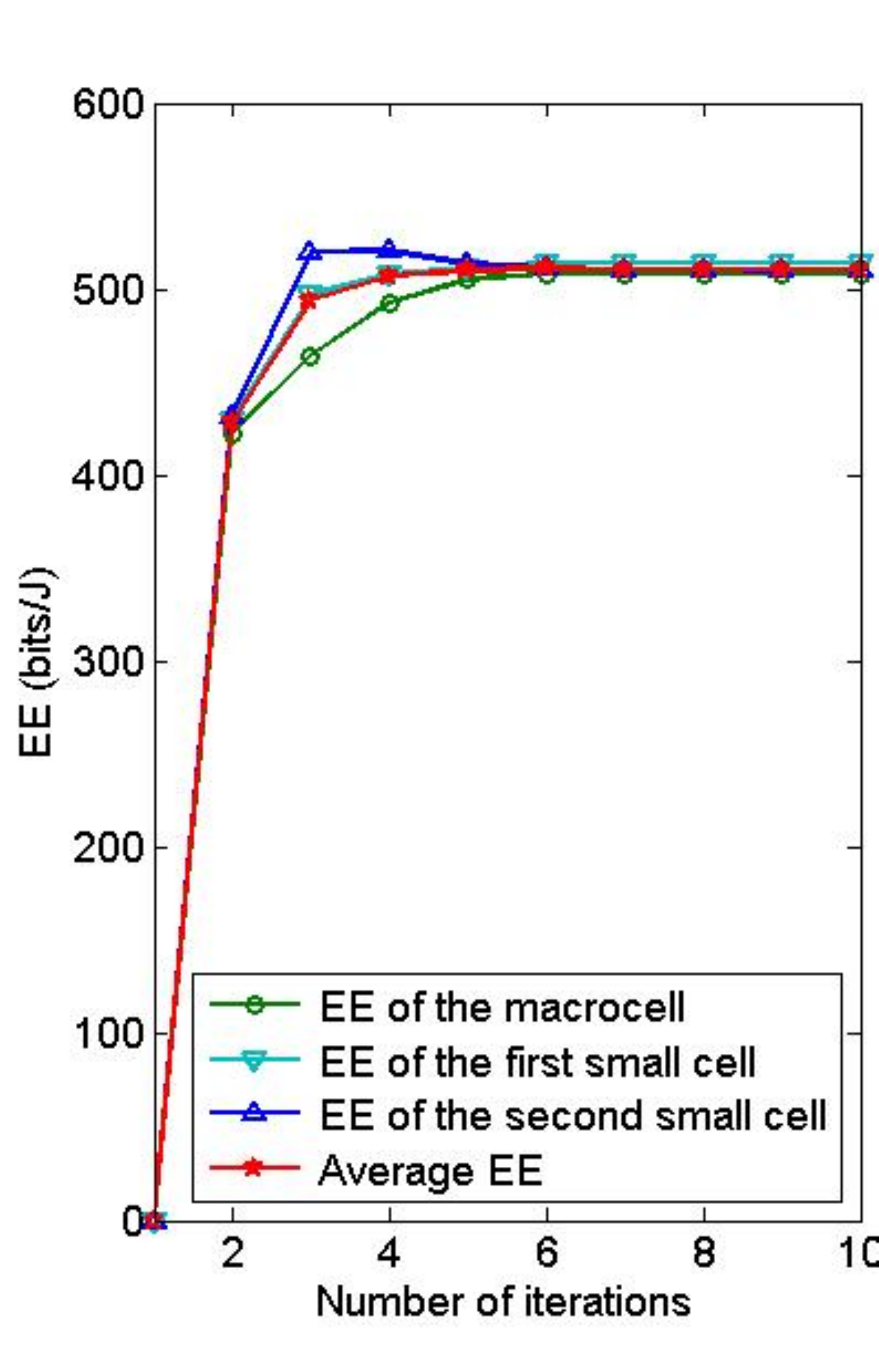}}
\subfigure[The NGT-based algorithm]{\label{2222}
\includegraphics[width=0.22\textwidth]{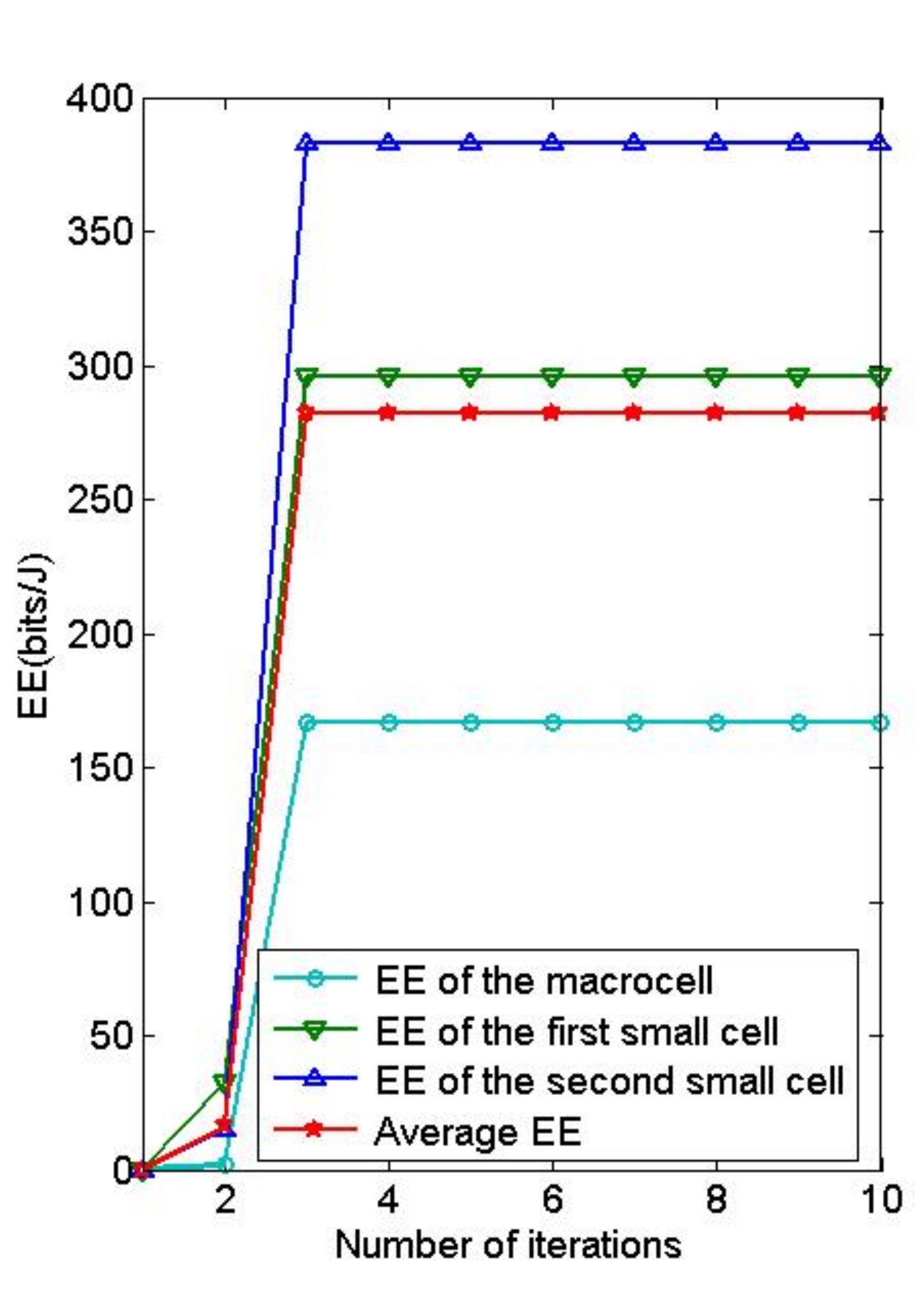}}
%\captionstyle{mystyle3}
\caption{EE versus the number of iterations with different power control algorithm.}
\label{fig2}
\end{figure}

In Fig. \ref{fig2}, we show the performance comparison between the proposed distributed algorithm and the NGT based algorithm in \cite{Liang}.
Fig. \ref{1111} depicts the EE versus the number of iterations with $N_u=6$ by using the proposed distributed algorithm.
Fig. \ref{2222} shows the EE versus the iterations with $N_u=6$ by using the NGT-based algorithm.
It can be observed that the EE of each cell approximately approaches the average EE by using the proposed distributed algorithm, while there are quite great differences among the cells by using the NGT-based algorithm.
It can be readily seen that the proposed distributed algorithm can greatly improve the fairness among the cells.

\begin{figure}[!t]
\centering %\vspace*{135pt}
\includegraphics[width=0.44\textwidth]{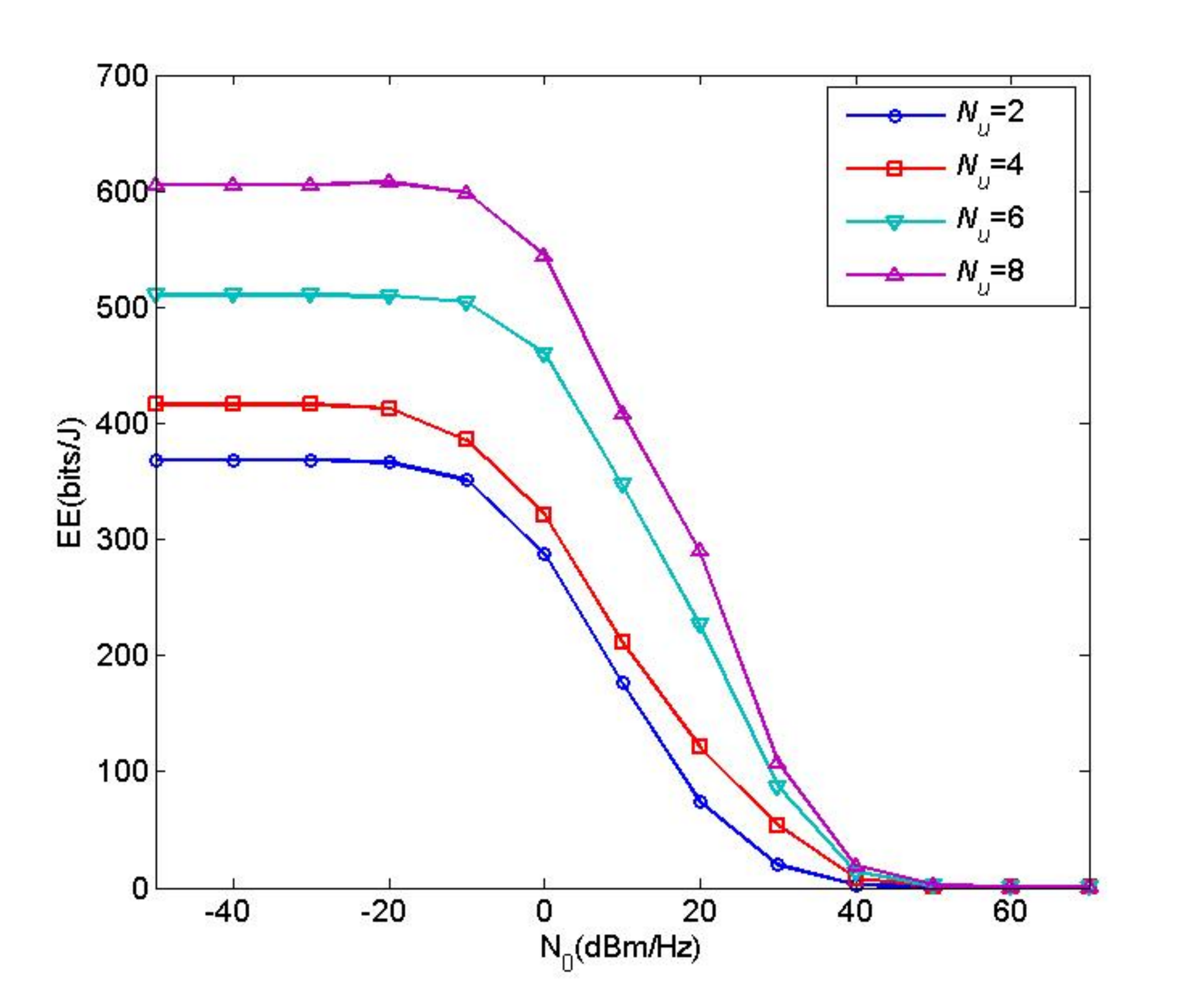}
%\captionstyle{mystyle3}
\caption{EE versus the noise with \(K=2\).}% and different $N_u$.}
\label{fig5}
\end{figure}

In Fig. \ref{fig5}, we show the EE versus the noise power $N_0$ with $K=2$ for different $N_u$.
It is obvious that the EE decreases with $N_0$ and the EE increases with $N_u$ by using the proposed EGT-based algorithm.

\section{Conclusions}
{In this paper, we have proposed a distributed energy efficient power control algorithm for the multi-user and multi-cell scenario in the uplink two-tier networks with small cells and massive MIMO.
On the one hand, the computational complexity has been greatly reduced from an exponential complexity to a linear complexity by using the proposed distributed power control algorithm.
On the other hand, the fairness among the cells has been remarkably improved by using EGT.
}

\section*{Acknowledgments}
This work was supported in part by the National Basic Research Program of China (973 Program 2012CB316004),
the National 863 Project (2015AA01A709),
and the Natural Science Foundation of China (61221002).

\bibliographystyle{IEEEtran}
%\bibliography{manuscript_smallcell_massiveMIMO}


%\newpage
\end{document}